\documentstyle[12pt,psfig,aasms4]{article}
\newcommand{\msun}{\mbox{${\rm M}_\odot$}}

\newcommand{\SeBa}{\mbox{\sf SeBa}}

\def\grbs{gamma-ray bursts}
\def\apgt{\ {\raise-.5ex\hbox{$\buildrel>\over\sim$}}\ }
\def\aplt{\ {\raise-.5ex\hbox{$\buildrel<\over\sim$}}\ }

\newcommand{\Meszaros}{\mbox{M\'esz\'aros}}
\slugcomment{Japan Society for the Promotion of Science Fellow}

\lefthead{S.\ F.\ Portegies Zwart}
\righthead{Gamma-ray binaries}

\begin{document}

\title{Gamma-ray binaries: stable mass transfer from neutron star to
       black hole} 
\author{Simon F.\ Portegies Zwart}
\affil{Dept.\ of General System Studies,
       University of Tokyo, 3-8-1 Komaba, Meguro-ku, Tokyo 153, Japan}
\authoremail{spz@grape.c.u-tokyo.ac.jp}

\begin{abstract}
Gamma-ray bursts are characterized by a duration of milliseconds
to several minutes in which an enormous amount of radiation is
emitted.

The origin of these phenomena is still unknown because proposed models
fail to explain all the observed features.  

Our proposed solution to this conundrum is a new class of
mass-exchanging binaries in which a neutron star transfers mass to a
black hole.  According to recent studies binaries which contain a
neutron star and a black hole are much more frequent than was
previously believed.  Mass exchange is driven by the emission of
gravitational waves but the redistribution of mass in the binary
system prevents coalescence.  The phase of mass transfer is
surprisingly stable and lasts for several thousands of orbital
revolutions (about a minute).  With a simple analytic model we
demonstrated that this new class of binaries could provide
an excellent candidate for the observed phenomena known as gamma-ray
bursts.
\end{abstract}


\section{Introduction}
The identification of the optical counterparts to gamma-ray bursts
GRB~970228\,(\cite{GGv97}) and GRB~970508\,
(\cite{He97}) and the measurement of a redshift of $z \geq
0.853$ (\cite{MDK97}) for the latter makes a cosmological origin for this
phenomenon hard to circumvent. If indeed cosmological, they must be
tremendously energetic: the maximum luminosity is many times 
the energy output of an entire galaxy and the total 
energy emitted is at least $10^{51}$ erg.

The rapid rise in luminosity and the short timescale variability
suggest that the radiation is generated in an area of only a couple of
hundred kilometers (a few light milliseconds).  The duration of the
burst (seconds to minutes) indicates that something within this region
is relatively stable. The complex temporal structure of the energy
release reflects the activity of a highly variable inner engine (Sari
\& Piran 1997; Kobayashi et al.\ 1997).

Suggestions have ranged from coalescing
neutron-star binaries\, (\cite{BNP84}), compact objects merging with the
central massive black hole of a galaxy\, (\cite{PFT94}) to
hypernovae\, (\cite{P98}). However all these models have great difficulty
explaining the duration (\cite{M97}) and the intrinsic
variability (Sari \& Piran 1997) of the burst.

Detailed studies of the stability of binaries where one neutron star
transfers mass to another neutron star reveal that coalescence is
inevitable within a few orbital revolutions owing the Darwin-Riemann
instability (\cite{CE77}; \cite{LRS93}).  Stable mass transfer can be
achieved only if the mass of the secondary is smaller than 67\% of the
primary\, (\cite{JK92}).  A neutron star which transfers mass to a black
hole with mass $\apgt 6$\,\msun\ cannot have a stable
orbit\, (\cite{LS76}).  Binaries of a neutron star and a lower mass black
hole have not been considered because a lack of observational
evidence.  Recent understanding of accretion on to a neutron star
(\cite{C93}; \cite{B95}; \cite{FBH96a}) has changed this view
completely, at least in the theoretical side. The birthrate (and also
the merger rate) of such systems might exceed the number of binary
pulsars by an order of magnitude (\cite{PZY98}; \cite{BB98}; Lipunov 1998).

\section{Formation and evolution} 
A binary consisting of two neutron stars is formed via a standard
scenario\, (\cite{HH72}) in which a stable phase of mass exchange is
followed by a supernova explosion, after which a common envelope phase
reduces the orbital separation (see scenario {\em I}\, in\, Portegies
Zwart \& Yungelson 1998 for a detailed description).  Recent
understanding of accretion on to a neutron star which is engulfed by
its companion in a common envelope allows the accretion rate to be
highly super Eddington (\cite{C93}; \cite{FBH96a}), as much as $10^8$
times larger or about 1\,\msun\ per year. The neutron star cannot
support this extra mass and collapses to a black hole.  The result is
a close binary system consisting of a neutron star and a black
hole. At the end of the common-envelope phase the mass of the black
hole is between 2.4\,\msun\ and 7.0\,\msun\ (\cite{WB96};
\cite{BB98}), the distribution within this mass interval is uncertain.

The separation between the two stars shrinks due to gravitational wave
radiation (see \cite{PM63}) until the neutron star
(with mass $m$) fills its Roche lobe. Mass transfer from the
neutron star to the black hole (with mass $M$) is still driven
by the emission of gravitational waves. The redistribution of mass in
the binary system, however, increases the separation and prevents
coalescence\, (\cite{KL97}). The time taken for material to travel
from the neutron star to the black hole is only a few milliseconds and
the neutrons do not have time to decay. The accretion rate is
therefore not Eddington limited and no mass is lost in the transfer
process; mass transfer proceeds conservatively, i.e.:
$\dot{M} = -\dot{m}$. This assumption is not that bold as in the first
few orbital revolutions the accretion stream passes the event horizon
of the black hole and the material falls in without forming an
accretion disc (see sect.\,\ref{sec:lightcurve}). The angular momentum
carried by this material is largely returned to the binary orbit
before it reaches the event horizon of the black hole.

The process is auto-regulated in that the increase in separation due
to the mass exchange over compensates the decrease in separation due
to the angular momentum loss, leading to a net increase in orbital
separation.

If the neutron star can be represented by the ideal neutron gas (a
Newtonian polytrope with index $n=3/2$) its radius $r$ is inversely
proportional to the cube root of its mass\, (\cite{A91}).  The
smallest possible mass of such a neutron star is around $0.1$\,\msun.
If the mass drops below this limit $\beta$ decay and nuclear fission
drive the explosion of the unstable neutron star\,
(\cite{CST93})\footnote{Using a realistic equation of state for the
neutron star shortens the binary lifetime but the main picture is
unchanged (C-H.\ Lee, private communication)}.

\begin{figure}[t]\vspace{0.3cm}
\centerline{\hspace{-0.8cm}
\psfig{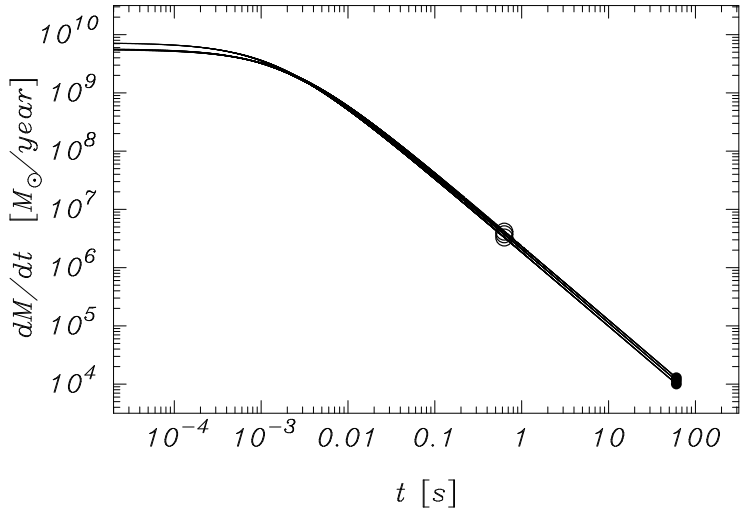}
}
\caption[]{
The mass transfer rate as a function of time for a
3\,\msun, 4\,\msun\ and a 5\,\msun\ black hole all accompanied by a
1.4\,\msun\ neutron star (the three lines are very similar).  The
$\circ$ indicates the moment when the accretion stream no longer
crosses the last stable orbit around the black hole. The $\bullet$ at
the right end of the curve indicates the moment that the mass of the
neutron star drops below 0.1\,\msun.  The phase of mass transfer for
the binary containing the 3\,\msun\ black hole lasted for
approximately 1.5 minutes, the higher mass binaries live shorter.
\label{Fig:mdot}
}
\end{figure}

The rate of mass transfer can be computed from the change in orbital
angular momentum $\dot{J}$ which has four components: $\dot{J} =
\dot{J}_{\rm orb} + \dot{J}_{\rm gw} + \dot{J}_{\rm bh} + \dot{J}_{\rm
ns}$, where $\dot{J}_{\rm orb}$ is given by the redistribution
of mass in the binary system, $\dot{J}_{\rm gw}$ gives the loss of
angular momentum due to the emission of gravitational waves
(\cite{PM63}, \cite{P64}) and $\dot{J}_{\rm bh}$ and $\dot{J}_{\rm
ns}$ are the variation in angular momentum for the rotation of the
black hole and neutron star, respectively. The latter has a negligible
contribution but $\dot{J}_{\rm bh}$ might be significant. For the
remainder of the discussion we assume both $\dot{J}_{\rm ns}$ and
$\dot{J}_{\rm bh}$ to be zero.  (The increase in orbital separation
found in the hydrodynamical simulations performed by Kalu\'zniak \& Lee
1997 is only about 25\% smaller than if conservation of mass and
angular momentum is assumed, apparently these assumption are not
seriously violated.)  Combining the equations with the requirement
that the donor fills its Roche-lobe $r_l$ (see \cite{P71} for a
approximate equation) and keeps doing this as mass is transferred,
i.e.: $\dot{r}_l = \dot{r}$, results in an expression for the required
rate of mass transfer
\begin{equation}
\dot{m} = {32 G^3 \over 5 \kappa^4 c^5} {m^{16/3} \over 
			         q^{2/3} (q+1)^{1/3} (q-2/3)}.
\label{Eq:mdot}\end{equation}
Here $q\equiv m/M$ and $G$, $c$ are the gravitational constant and the
speed of light. The constant $\kappa \approx 2.2 \msun^{1/3}r_{\odot}
\approx 4.1 \times 10^{17}$\,[cm\, g$^{1/3}$], with $r_{\odot}$ the
radius of a 1\,\msun\ neutron star. Figure\,\ref{Fig:mdot} presents
the mass accretion rate as a function of time.


We computed the birth rate of such binaries with the detailed binary
population synthesis program \SeBa\, (\cite{PZV96}; model H
of\,\cite{PZY98}). It is $10^{-4.3}$\,per year in the Galaxy. In
comparison, the birthrate of double neutron star systems is
$10^{-5.2}$\,per year. The majority of these systems begin mass
transfer within a billion years after formation resulting in a merger
rate of $\sim 10^{-4.5}$\,per year.  Bethe \& Brown\, (1998)
independently compute birth- and merger rates and obtain similar
results. A somewhat smaller rate is derived from the observed
population of neutron star binaries by Phinney (1991) and Narayan et
al.\ (1991).

\section{Stability}
Gravitational wave radiation circularizes the orbit of the binary and
let the separation shrink.  The spiral in owing to the emission of
gravitational waves can be arrested by mass transfer from the neutron
star to the black hole if the mass ratio is less than $2/3$
(Eq.\,\ref{Eq:mdot}, see also \cite{JK92}; \cite{K92b}).  Two neutron
stars in observed binary pulsars have almost the same mass and
coalescence is expected to occur within a few orbital periods upon
Roche-lobe contact.  To prevent immediate coalescence the mass of the
accretor must exceed 2.1\,\msun\, if the donor is a Chandrasekhar mass
neutron star (1.4\,\msun).

The orbital separation at Roche-lobe contact must exceed the last
stable orbit, i.e, $a > 3 R_{\rm Sch}$ where $R_{\rm Sch} \equiv
2GM/c^2$ is the Schwarzschild radius. For a Kerr black hole, which
is more appropriate for our discussion, the stability limit is even
more relaxed.


\begin{figure}[t]\vspace{0.3cm}
\centerline{\hspace{-0.8cm}
\psfig{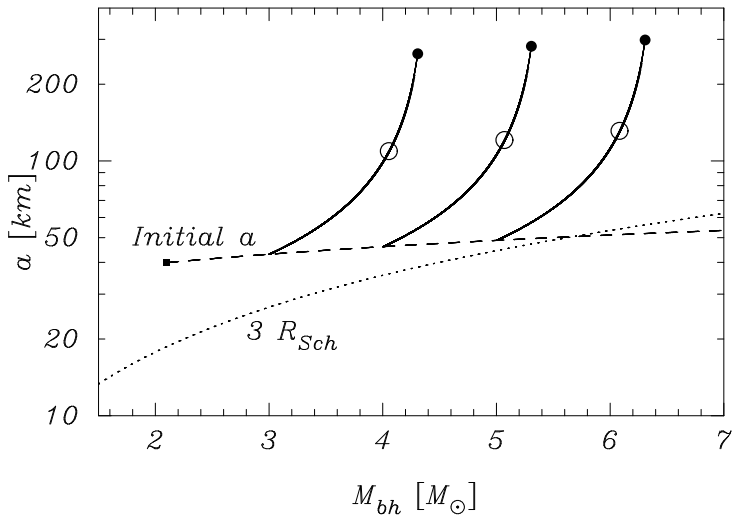}
}
\caption[]{
Evolutionary tracks for a 3\,\msun\ black hole, a
4\,\msun\ and a 5\,\msun\ black hole accompanied by a 1.4\,\msun\
neutron star through the phase of mass transfer (left, middle and right solid
lines, respectively). The mass of the black hole is on the horizontal
axis the vertical axis gives the orbital separation in kilometers. The
evolution ends at the $\bullet$ when the neutron star becomes
unstable. The dotted line gives the separation below which the binary
is gravitationally unstable ($3R_{\rm Sch}$). The dashed line gives
the initial orbital separation at the moment mass transfer starts.
The initial and also minimum orbital separation (dashed line) is
larger than $3 R_{\rm Sch}$ for black holes with a mass smaller than
about 5.8\,\msun. The square to the left of the dashed line indicates
the minimum mass for the accretor in order to have stable mass
transfer from a 1.4\,\msun\ neutron star. \\ The $\circ$\, indicates
the moment in the evolution of the binary at which a disc around the
black hole can be formed. From that moment the accretion stream
remains well outside the last stable orbit.
\label{Fig:sma_ev}}
\end{figure}

At the onset of mass transfer the horizon radius of the black hole
is a considerable fraction of its Roche lobe and the accretion
stream falls in practically radially: there will be no accretion disc.
Once the orbital separation has increased sufficiently a neutron rich
disc can form around the black hole.  Figure\,\ref{Fig:sma_ev}
illustrates the evolution of such binaries and the stability criterion.

The time scale for orbital decay by gravitational radiation
is shorter than the time scale for tidal synchronization
([\cite{BC92}; \cite{LRS94}], note however that the liquefaction of the
neutron star just before Roche-lobe overflow [\cite{K92b}], speeds up
synchronization [\cite{LRS93}]) so that the binary is not tidally
synchronized at the moment mass transfer begins and the Roche-geometry
is strictly speaking not applicable.  What effect this has
on the computation is unknown. It possibly increases the mass
transfer rate and decreases the systems' lifetime, but the binary does
{\em not} become tidally unstable\, (\cite{KL97}).

\section{Gamma-ray production}\label{sec:lightcurve}

The energy available from the infall of material into the potential
well of the black hole, $L \propto GM\dot{M}/R_{\rm Sch}$, is not
likely to drive the gamma ray burst, because the transformation of
this energy into gamma-rays is not efficient enough
(Shapiro 1973; Shrader \& Titarchuk 1998).

How to get the energy out in the form of gamma-rays is not clear. An
interesting model is based on the Blandford-Znajek (1977) mechanism
where the rotation of a rapidly spinning Kerr black hole is used as an
energy source (see e.g.\ \Meszaros\ \& Rees 1997 and Katz 1997 for 
details). The rotational energy of such a black hole is
approximately $10^{54}$ erg, but the fraction which is liberated is
considerably smaller (Macdonald et al.\ 1986):
\begin{equation}
	L \approx 10^{50} \left( {\mu M \over 3 [\msun]} \right)^2
	                  \left( {B \over 10^{15} [G]} \right)^2 
	  \;\; [{\rm erg\,s}^{-1}].
\end{equation}
Here $\mu$ is the angular momentum of the black hole relative to that
if maximally rotating.  An enormous magnetic field $B$ is required and
how it is generated is not well understood. However, strong
magnetic fields in black holes have gained a lot of support over the
last few years (see e.g.\ Paczy\'nski 1998 for an overview).  The
strong magnetic field is anchored in the disc but the power comes from
the spin energy of the black hole (Kats 1997).

The magnetic field causes the radiation to be collimated along the
axis of the black hole. If the opening angle of the emission is
limited to $\aplt 10^\circ$ the efficiency of the radiation process
can be small $\aplt 10^{-4}$ and still produce a phenomenon energetic
enough to power a high redshift gamma-ray burst (\Meszaros\ \& Rees
1997). Such a small opening angle also conveniently increases the low
occurrence rate of gamma-ray bursts of about $10^{-7}\; {\rm Mpc}^{-3}\;
{\rm yr}^{-1}$ to the rate of mergers between compact objects as given
earlier.

\section{Baryon pollution}
The gamma-ray production would be greatly reduced if too many baryons
(which absorb the fotons) would pollute the vicinity of the black
hole. Basically there are two sources of baryons; the inner
edge of the accretion disc, and the baryon loaded wind from the
neutron star surface.

In the presented model a disc is formed only after a phase of radial
infall in the black hole, and little or no baryon pollution is
expected in this phase. After the formation of a disc (at which point
the accretion rate has dropped by three orders of magnitude) the
region near the axis of the black hole is still expected to be
reasonable free of
material; the black hole easily accretes material with angular
momentum below a specific value (Fishbone \& Moncrief 1976, see 
however Chakrabarti 1998 for counter arguments).
Whether or not the vicinity of the black hole is clean enough to allow for
the high Lorentz factors required to produce the gamma-ray burst is
unclear. 

The neutron star is heated by the tidal friction as it is forced into
co-rotation (see Kochanek 1992). This may result in a baryon loaded
wind from the neutron star surface which pollutes the environment. The
contamination of baryons due to this tidal heating is computed by
\Meszaros\ \& Rees (1992).  In the presented model the binary spirals
outwards instead of inwards, and the heat production due to tidal
friction decreases in time.

\section{Discussion}

The time structure of gamma ray bursts must reflect the time
structure of their energy release (Sari \& Piran 1997).  
About three quarters of the observed \grbs\ have a duration of seconds
to several minutes\, (\cite{MFB98}). Fireball models can only explain
this duration with an enormous time dilation (\cite{RM92}) and are
unable to explain the temporal structure. 

The shortest bursts can still be understood as the merging of two
neutron stars or as the cases where mass transfer from a neutron star
to a black hole is unstable. The tail of the
light curve is not necessarily detected in its full extent because the
luminosity of the burst drops below detector sensitivity. The presented
model therefore explains a range of timescales from milliseconds to
minutes and predicts a relative frequency of at least 1 short burst to
17 long bursts (given by the relative frequency of merging neutron
stars and mass transferring neutron star--black hole systems\,
(\cite{PZY98}).  In fact 3 out of ten observed bursts are short\,
(\cite{MFB98}).

In the accretion phase our class of objects is very similar to that of
low-mass X-ray binaries with an accreting black hole, and the active
cores of galactic nuclei (Protheroe \& Kazanas 1983).  The X-ray
binaries are highly variable (\cite{ZWH94}; \cite{BMK97}) and the
detection of gamma rays have been reported (\cite{GJK98}).  

When the neutron star reaches its lower-mass limit it explodes
generating a luminosity of $10^{49}$ to $10^{51}$\,erg/s accompanied
by a burst of anti-neutrinos of $10^{51}$ to $10^{52}$ erg/s
(\cite{CST93}).  By this time the energy emitted in gamma-rays has
probably stopped or decreased below detector sensitivity. The radius
of the neutron star at that moment (at a mass of about 0.1\,\msun) is
about 33\,km, and the orbital separation is more than 250\,km. The
escape velocity from the black hole at this distance is less than
$10^5$\,km/s. With an expansion velocity for the exploding neutron
star of several $10^4$\,km/s (\cite{CST93}) only a small fraction of
its mass can escape, so the black hole is released with a velocity of
only a few tens of kilometers per second.

With a birth rate of $10^{-4.5}$\,yr$^{-1}$ (\cite{PZY98}) our Galaxy
has produced about $10^{5.5}$ low velocity black holes with a mass
between 4\,\msun\ and 8\,\msun\ over the last 10\,Gyr (for a
constant star formation rate).  Per event about $10^{-3}$\,\msun\ of
neutron star material is ejected enriching the Galaxy with a total of
$10^3$\,\msun\ in r-processed elements so that the majority of the
Galactic thorium enrichment could have originated in \grbs. This
enrichment is about two orders of magnitudes larger than previous
estimates based on mergers between neutron stars\, (\cite{R97}).

\section{Conclusion}
The recent understanding of the super Eddington accretion process on
to a neutron star in a common-envelope phase predicts the formation of
a large number of close binaries in which a neutron star is
accompanied by a low mass ($\aplt 7$\,\msun) black hole.
Model computations predict that the formation rate of such binaries
exceeds the formation of classical high-mass binary pulsars by about
an order of magnitude. These binaries are therefore expected to
contribute appreciably to the rate of any associated observable
phenomena.

There are three possible forms of mass transfer from a neutron star to
its companion: unstable mass transfer to another neutron star and
stable and unstable transfer to a black hole.  Also for gamma-ray
bursts there is evidence for three types: short and faint bursts, long
and bright and a third class of intermediate bursts\,(\cite{MFB98}).

In the stable binaries the duration of mass transfer lasts for about a
minute after which the neutron star explodes.  An accretion disc with
a mass $\aplt 0.4$\,\msun\ around the black hole develops about a
second after the onset. The gamma-ray burst is powered by using
the rotational energy of the black hole via the Blandford-Znajek
(1977) mechanism. If the neutron star finally explodes it is
accompanied by a burst of luminosity as well as anti neutrinos. The
accompanying black hole is released with a low velocity.  As the
binary spirals outwards instead of inwards a reversed chirp in the
gravitational wave signal is expected, the signal-to-noise, however,
drops considerably during this process.

In our model gamma-ray bursts are another exciting class of
mass-exchanging binary stars.  We propose to call them gamma-ray
binaries.

\acknowledgments I am grateful to Gerald Brown, Junichiro Makino,
Gijs Nelemans and Chirstopher Tout for discussions and checking
computations as well as English. 

\clearpage

\end{document}